\documentclass[aps,prc,twocolumn,tightenlines,superscriptaddress,a4paper]{revtex4}
\usepackage{dcolumn}
\usepackage{graphicx}
\usepackage{rotating}
\usepackage{natbib}
\usepackage{amsmath}
\usepackage{braket}

\begin{document}

\title{One- and two-neutron removal reactions from the most neutron-rich carbon isotopes}

\author{N.~Kobayashi}
    \affiliation{Department of Physics, Tokyo Institute of Technology,
    2-12-1 O-Okayama, Meguro, Tokyo 152-8551, Japan}
\author{T.~Nakamura}
    \affiliation{Department of Physics, Tokyo Institute of Technology,
    2-12-1 O-Okayama, Meguro, Tokyo 152-8551, Japan}
\author{J.~A.~Tostevin}
    \affiliation{Department of Physics, Tokyo Institute of Technology,
    2-12-1 O-Okayama, Meguro, Tokyo 152-8551, Japan}
    \affiliation{Department of Physics, Faculty of Engineering and
      Physical Sciences, University of Surrey, Guildford,
      Surrey GU2 7XH, United Kingdom}
\author{Y.~Kondo}
    \affiliation{Department of Physics, Tokyo Institute of Technology,
    2-12-1 O-Okayama, Meguro, Tokyo 152-8551, Japan}
\author{N.~Aoi}
    \affiliation{RIKEN Nishina Center, Hirosawa 2-1, Wako,
    Saitama 351-0198, Japan}
\author{H.~Baba}
    \affiliation{RIKEN Nishina Center, Hirosawa 2-1, Wako,
    Saitama 351-0198, Japan}
\author{S.~Deguchi}
    \affiliation{Department of Physics, Tokyo Institute of Technology,
    2-12-1 O-Okayama, Meguro, Tokyo 152-8551, Japan}
\author{J.~Gibelin}
    \affiliation{LPC-Caen, ENSICAEN, IN2P3-CNRS, Universit$\acute{e}$ de Caen,
    14050, Caen Cedex, France}
\author{M.~Ishihara}
    \affiliation{RIKEN Nishina Center, Hirosawa 2-1, Wako,
    Saitama 351-0198, Japan}
\author{Y.~Kawada}
    \affiliation{Department of Physics, Tokyo Institute of Technology,
    2-12-1 O-Okayama, Meguro, Tokyo 152-8551, Japan}
\author{T.~Kubo}
    \affiliation{RIKEN Nishina Center, Hirosawa 2-1, Wako,
    Saitama 351-0198, Japan}
\author{T.~Motobayashi}
    \affiliation{RIKEN Nishina Center, Hirosawa 2-1, Wako,
    Saitama 351-0198, Japan}
\author{T.~Ohnishi}
    \affiliation{RIKEN Nishina Center, Hirosawa 2-1, Wako,
    Saitama 351-0198, Japan}
\author{N.~A.~Orr}
    \affiliation{LPC-Caen, ENSICAEN, IN2P3-CNRS, Universit$\acute{e}$ de Caen,
    14050, Caen Cedex, France}
\author{H.~Otsu}
    \affiliation{RIKEN Nishina Center, Hirosawa 2-1, Wako,
    Saitama 351-0198, Japan}
\author{H.~Sakurai}
    \affiliation{RIKEN Nishina Center, Hirosawa 2-1, Wako,
    Saitama 351-0198, Japan}
\author{Y.~Satou}
    \affiliation{Department of Physics and Astronomy,
    Seoul National University, 599 Gwanak, Seoul 151-742, Korea}
\author{E.~C.~Simpson}
    \affiliation{Department of Physics, Faculty of Engineering and
      Physical Sciences, University of Surrey, Guildford,
      Surrey GU2 7XH, United Kingdom}
\author{T.~Sumikama}
    \affiliation{Department of physics, Tokyo University of Science,
    Chiba 278-8510, Japan}
\author{H.~Takeda}
    \affiliation{RIKEN Nishina Center, Hirosawa 2-1, Wako,
    Saitama 351-0198, Japan}
\author{M.~Takechi}
    \affiliation{RIKEN Nishina Center, Hirosawa 2-1, Wako,
    Saitama 351-0198, Japan}
\author{S.~Takeuchi}
    \affiliation{RIKEN Nishina Center, Hirosawa 2-1, Wako,
    Saitama 351-0198, Japan}
\author{K.~N.~Tanaka}
    \affiliation{Department of Physics, Tokyo Institute of Technology,
    2-12-1 O-Okayama, Meguro, Tokyo 152-8551, Japan}
\author{N.~Tanaka}
    \affiliation{Department of Physics, Tokyo Institute of Technology,
    2-12-1 O-Okayama, Meguro, Tokyo 152-8551, Japan}
\author{Y.~Togano}
    \affiliation{RIKEN Nishina Center, Hirosawa 2-1, Wako,
    Saitama 351-0198, Japan}
\author{K.~Yoneda}
    \affiliation{RIKEN Nishina Center, Hirosawa 2-1, Wako,
    Saitama 351-0198, Japan}
\date{\today}

\begin{abstract}
The structure of $^{19,20,22}$C has been investigated using
high-energy (about 240~MeV/nucleon) one- and two-neutron removal
reactions on a carbon target. Measurements were made of the inclusive
cross sections and momentum distributions for the charged
residues. Narrow momentum distributions were observed for one-neutron
removal from $^{19}$C and $^{20}$C and two-neutron removal from
$^{22}$C. Two-neutron removal from $^{20}$C resulted in a relatively
broad momentum distribution. The results are compared with
eikonal-model calculations combined with shell-model structure
information. The neutron-removal cross sections and associated
momentum distributions are calculated for transitions to both the
particle-bound and particle-unbound final states. The calculations
take into account the population of the mass $A-1$ reaction residues,
$^{A-1}$C, and, following one-neutron emission after one-neutron
removal, the mass $A-2$ two-neutron removal residues, $^{A-2}$C. The
smaller contributions of direct two-neutron removal, that populate the
$^{A-2}$C residues in a single step, are also computed. The data and
calculations are shown to be in good overall agreement and consistent
with the predicted shell-model ground state configurations and the
one-neutron overlaps with low-lying states in $^{18-21}$C. These
suggest significant $\nu{s}_{1/2}^2$ valence neutron configurations in
both $^{20}$C and $^{22}$C. The results for $^{22}$C strongly support
the picture of $^{22}$C as a two-neutron halo nucleus with a dominant
$\nu{s}_{1/2}^2$ ground state configuration.
\end{abstract}

\maketitle

\section{Introduction \label{intro}}
Residue momentum distributions following the dissociation of halo
nuclei have long been recognized as probes of the spatially extended
valence neutron wave functions \cite{Kob88,Orr92}. Over the last
decade, high-energy nucleon removal reactions have become a powerful
spectroscopic tool to explore the structure of nuclei far from
stability. Specifically, the momentum distributions and associated
cross sections offer a means to probe the active single-particle
orbitals near the Fermi surface, whereby the shapes of the momentum
distributions reflect the orbital angular momentum of the removed
nucleon(s) and the cross sections the spectroscopic strength
\cite{Han96,BaV98,HaT03,BeH04,Gad08}.

Two-nucleon (2$N$) removal reactions are challenging experimentally
and remain less well studied \cite{Baz03,Tos04,Yon06,Tos06}. Recently,
it has been clarified that 2$N$ removal cross sections and the
associated residual nucleus momentum distributions, following direct
removal of two well-bound like nucleons, can provide a sensitive probe
of rapid structural changes \cite{Gadxx,Fal}, of spins of final-states
\cite{Tos07,simxx}, and also of aspects of two-nucleon spatial
correlations near the nuclear surface \cite{Tos06,simzz}. However, a
complication arises when discussing reactions that result in the
removal of two weakly-bound (valence) neutrons from a neutron-rich,
mass $A$, projectile. The mass $A-2$ reaction residues can now result
from two distinct mechanisms: single-step direct knockout of a pair of
neutrons and one-neutron knockout followed by neutron emission from
excited particle-unbound states of the intermediate mass $A-1$
residue. The latter will be populated with cross sections that are
typically an order of magnitude or more larger than those for the
direct pair removal. Both processes are discussed quantitatively in
the analyses presented here.

This article reports the first measurements of one and two-neutron
removal from $^{19,20,22}$C secondary beams at about
240~MeV/nucleon. The carbon isotopic chain is of considerable interest
from a structural point of view, as it exhibits large odd-even
staggering in the one-neutron separation energies (see e.g., Fig.~2 of
\cite{ECS_C}), and weak $s$-wave valence neutron binding and halo
formation for both $^{15}$C \cite{Sau00}, $S_{1n}$ = 1.22(1) MeV, and
$^{19}$C \cite{Naka19}, $S_{1n}$ = 0.58(9) MeV. The next heavier
odd-$A$ halo candidate, $^{21}$C, is known to be particle unbound
while $^{22}$C is bound by $S_{2n} = 0.42(94)$~MeV, which is estimated
on the basis of systematics \cite{Aud03}. As such, $^{22}$C may be
pictured as a three-body ($^{20}$C + $n$ + $n$) Borromean system --
having no bound two-body subsystems. The structures of the $^{19-22}$C
ground and excited states are considered here by the use of the fast
neutron removal reaction methodology.

A recent measurement of the reaction cross section of $^{22}$C on a
proton target \cite{Tan10} has also drawn much attention. Its
observation of an enhanced cross section suggests that $^{22}$C has
(a) an extended matter density and (b) a two-neutron halo ground state
structure, dominated by a weakly-bound $s$-wave two-neutron
configuration. Such measurements, carried out over a wide energy
range, have been successful in deducing effective nuclear matter radii
and their evolution along isotopic chains. However, they cannot
provide specific information about emerging changes in the microscopic
nuclear structures.

Fast one- and two-neutron removal reactions from the lighter
neutron-rich carbon isotopes, up to $^{19}$C, have been studied
experimentally using beryllium and carbon targets (see e.g.,
\cite{Sau00,Baz98,Oza08,Mad01,Yam03,Sau04,Ter04,Chi04,Wu05}). Results
for neutron removal from $^{18,19}$C on a proton target were reported
and analyzed by Kondo {\em et al}. \cite{Kondo}.  The incident beam
energies of these studies ranged from 45$-$103~MeV/nucleon. In several
cases the parallel or transverse momentum distributions of the
reaction residues were also measured. More recently, Ozawa {\em et
al.} \cite{Oza11} measured one- and two-neutron removal reactions of
$^{19,20}$C using a proton target at 40 $A$ MeV. However, it was
difficult to discuss the reactions for $^{20}$C due to the low
statictics. A systematic analysis of many of these results, using the
shell model and reaction formalism exploited here, was presented in
Ref.\ \cite{ECS_C}.

As mentioned above, the measurements reported here were made at about
240~MeV/nucleon. At such an energy domain the underlying assumptions
of the sudden and eikonal approximations for the reactions are well
founded. This paper thus presents a quantitative analysis of the
results of both (a) direct one- and two-neutron removal, and (b)
indirect two-neutron removal. In doing so we are able to elucidate the
dominant single-particle structure of $^{19,20,22}$C and the extent to
which these are tracked by shell-model calculations at the boundary of
the $p$-shell and the lower part of the $sd$-shell. Only the
shell-model currently provides this level of information on the
nucleon single-particle structures consistently along an isotopic
chain in a form that can be readily used in reaction calculations.
Thus we employ a consistent shell-model and reaction theory
description for all of the isotopes considered.

In Section \ref{expt} we discuss the experimental techniques. The
results are described in Section \ref{exp_results}. The theoretical
approach used is outlined in Section \ref{framework}. There both the
reaction theory and the shell-model spectroscopic strengths and
description of the neutron bound states needed for the calculations
are discussed. The experimental results and calculations are compared
and discussed in Section \ref{results}. The paper then concludes with
a brief summary.

\section{Experimental setup and details \label{expt}}

\begin{figure}[tp]
\begin{center} \includegraphics[width=0.95\columnwidth]{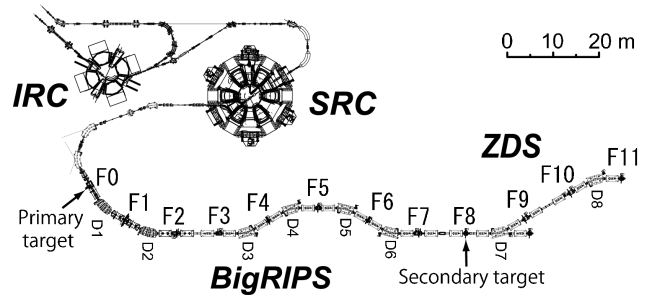}
\caption{A schematic view of the layout of BigRIPS (F0-F8) and the
zero degree spectrometer, ZDS (F8-F11), where F1 through F11 represent
each focal plane. The secondary carbon target was mounted at the
achromatic focal plane F8. The $^{18,19,20}$C residues were identified
using the ZDS.}
\label{ribf}
\end{center}
\end{figure}

\begin{figure}[tbp]
\begin{center} \includegraphics[width=0.95\columnwidth]{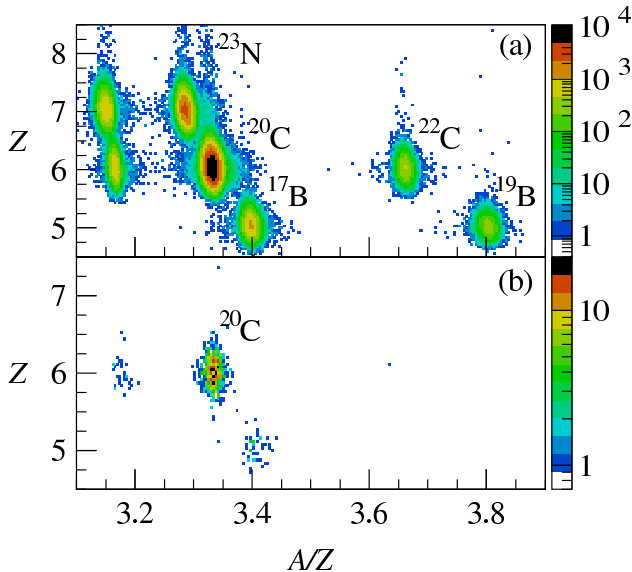}
\caption{(color online). Particle-identification spectrum of (a) the
$^{22}$C beam by BigRIPS, and (b) the residues in the ZDS after
selecting $^{22}$C ions before the secondary target.}
\label{pid}
\end{center}
\end{figure}

The experiments were performed at the RI Beam factory (RIBF) operated
by the RIKEN Nishina Center and the Center for Nuclear Study,
University of Tokyo. Secondary beams of $^{19,20,22}$C were produced
by fragmentation of a $^{48}$Ca beam at 345~MeV/nucleon on a
20-mm-thick rotating Be target. The typical $^{48}$Ca primary beam
intensities ($I_1$) for each setting are listed in Table
\ref{tbl:beams}. The secondary beams were separated using the
superconducting separator BigRIPS \cite{Kubo,Ohnishi} whose layout is
shown in Fig.~\ref{ribf}. An achromatic aluminum energy degrader of
15-mm-median thickness was located at the dispersive focal plane
F1. For the setting to produce the $^{19}$C beam, an aluminum degrader
of 8-mm-median thickness was also installed at the dispersive focal
plane F5. The secondary beam particles were identified event-by-event
by combining the measured time-of-flight (TOF), energy loss ($\Delta
E$), and magnetic rigidity ($B\rho$). The TOF was recorded between two
plastic scintillators at the achromatic focal planes F3 and F7,
$\Delta E$ was measured using a plastic scintillator at F7, and
$B\rho$ was determined from a position measurement using a plastic
scintillator read out on both sides by photomultiplier tubes at F5. An
example of the particle-identification spectrum is shown in
Fig.~\ref{pid}(a) in terms of the atomic number ($Z$) and
mass-to-charge ratio ($A/Z$), which demonstrates the clear separation
of $^{22}$C. The secondary beam intensities ($I_2$) and the momentum
spread ($\Delta P/P$) of the $^{19,20,22}$C beams are listed in Table
\ref{tbl:beams}. A secondary carbon target (4.02~g/cm$^2$) was mounted
at the achromatic focal plane F8. The mid-target energies
($\bar{E}/A$) of $^{19,20,22}$C were 243, 241, and 240~MeV/nucleon,
respectively, as listed in Table \ref{tbl:expcs}.

\begin{table}[bp]
\caption{The typical $^{48}$Ca primary beam intensity ($I_1$) for each
setting and typical secondary beam intensity ($I_2$). Also tabulated
are the momentum spread ($\Delta P/P$) of the secondary beams.}

\label{tbl:beams}
\begin{center}
\begin{ruledtabular}
\begin{tabular}{cccc}
Secondary beam & $I_1$ (pnA) & $I_2$ (particles/s) & $\Delta P/P$ \\ \hline
\rule [0pt]{0pt}{10pt}$^{19}$C    & $\approx$4  & $\approx$${1} \times 10^3$ & $\pm 2\%$ \\
$^{20}$C  & $\approx$6  & $\approx$${6} \times 10^2$ & $\pm 3\%$ \\
$^{22}$C   & $\approx$80 & $\approx$6               & $\pm 3\%$ \\
\end{tabular}
\end{ruledtabular}
\end{center}
\end{table}

The $^{18,19,20}$C residues following the reaction were collected by
tuning the rigidity of the zero degree spectrometer (ZDS) to center
the momentum distribution. The residues as well as the secondary beam
particles were identified event-by-event by combining TOF, $\Delta E$,
and $B\rho$, where TOF was measured between two plastic scintillators
at the achromatic focal planes F7 and F11 and $\Delta E$ was measured
using an ionization chamber at F11. The $B\rho$ was determined from a
position measurement using PPACs at the dispersive focal plane F9. The
resulting particle-identification spectrum for the $^{20}$C residues
is shown in Fig.~\ref{pid}(b).

The present study measured one- and two-neutron removal cross sections
($\sigma_{-xn}^{\rm exp}$, $x$ = 1 or 2) and the residue parallel
momentum ($P_{||}$) distributions for the channels listed in Table
\ref{tbl:expcs}. The cross sections were derived from the numbers of
the beam particles counted before the secondary target and those of
the residues registered at the final focal plane F11 of the ZDS, using

\begin{equation}
\sigma_{-xn}^{\rm exp} = \left(\frac{N_i^\prime}{N_i} - \frac{N_o^\prime}{N_o}
\right) \left( \frac{\sigma_R-\sigma_R^\prime}{e^{-\sigma_R^\prime
N_t}-e^{-\sigma_R N_t}} \right)\ .
\label{eqn:cross_section}
\end{equation}
Here $N_i$ ($N_o$) represents the number of projectiles and
$N_i^\prime$ ($N_o^\prime$) the number of residues for target-in ($i$)
(target-out ($o$)) runs. $N_t$ and $\sigma_R$ ($\sigma_R^\prime$) are
the number of target nuclei per unit area and the reaction cross
section of the projectile (residue). The background events were
subtracted by using target-out runs. Owing to the substantial
thickness of the carbon target, correction had to be applied to
account for the losses in the number of projectiles and residues owing
to reactions in the target. To do so, total reaction cross sections of
1375~mb for $^{22}$C, 1090~mb for $^{20}$C, 1139~mb for $^{19}$C, and
1023~mb for $^{18}$C, which were estimated using eikonal calculations,
were employed. It should be noted that a 10\% deviation in these
reaction cross section values translates into a (essentially
negligible) 1\% deviation of the deduced one- and two-neutron removal
cross sections. To obtain the residue momentum in the projectile rest
frame, the Lorentz transformation was applied to the laboratory system
residue momentum extracted from the TOF between F8 and F11. The
velocity of the projectile was obtained from the TOF between F3 and
F7. The cross sections $\sigma_{-xn}^{\rm exp}$ and the parallel
momentum $P_{||}$ distributions were corrected for the transmission
efficiency through the ZDS. These were estimated using a Monte Carlo
simulation, an angular acceptance of the ZDS obtained by calibration
runs using the secondary beams, and the design value of the momentum
acceptance ($\Delta P/P \leq \pm 3\%$) of the ZDS. To obtain a higher
transmission the momentum acceptance of BigRIPS was restricted to
$\Delta P/P \leq \pm 2\%$ in the analysis. The transmission was about
$90\%$ in all cases.

\section{Experimental results\label{exp_results}}

The one- and two-neutron removal cross sections extracted here are
summarized in Table \ref{tbl:expcs} together with the corresponding
mid-target energy of the projectile. The cross section for C($^{19}$C,
$^{18}$C) is smaller than that measured at lower energies. Near
60~MeV/nucleon for instance \cite{Chi04}, $\sigma_{-1n}$ = 226(65)
mb. In part this reflects the smaller intrinsic nucleon-nucleon cross
sections at the energy of the current experiment (being closer to the
minimum near 300~MeV) as well as changes in the real parts of the
optical potentials and a reduced diffractive breakup contribution for
energies in excess of 100 MeV/nucleon \cite{HBE}.  For $^{20}$C, the
yield of $^{19}$C residues is smaller than that for $^{18}$C,
reflecting the available energy windows for final states below the
first neutron thresholds in $^{18}$C ($S_n=4.18$~MeV) and $^{19}$C
($S_n=0.58(9)$~MeV). For $^{22}$C we note an enhanced cross section
compared to $^{20}$C which, as we will show, reflects the two-neutron
halo character of $^{22}$C.

\begin{table}[bp]
\caption{ The one- and two-neutron removal cross sections for each
reaction channel at the mid-target energies $(\bar{E}/A)$ shown. The
widths (FWHM) of the momentum distributions, after unfolding the
experimental resolutions, are also shown.}

\label{tbl:expcs}
\begin{center}
\begin{ruledtabular}
\begin{tabular}{lcccc}
Reaction & $\bar{E}/A$ (MeV) & $\sigma_{-xn}^{\rm exp}$ (mb) & FWHM (MeV/$c$) \\ \hline
\rule [0pt]{0pt}{10pt}C($^{19}$C, $^{18}$C) & 243 & 163(12) & 73(5) \\
C($^{20}$C, $^{19}$C) & 241 & 58(5)   & 46(13) \\
C($^{20}$C, $^{18}$C) & 241 & 155(25) & 204(11)\\
C($^{22}$C, $^{20}$C) & 240 & 266(19) & 76(8) \\

\end{tabular}
\end{ruledtabular}
\end{center}
\end{table}

\begin{figure}[tbp]
\begin{center} \includegraphics[width=0.95\columnwidth]{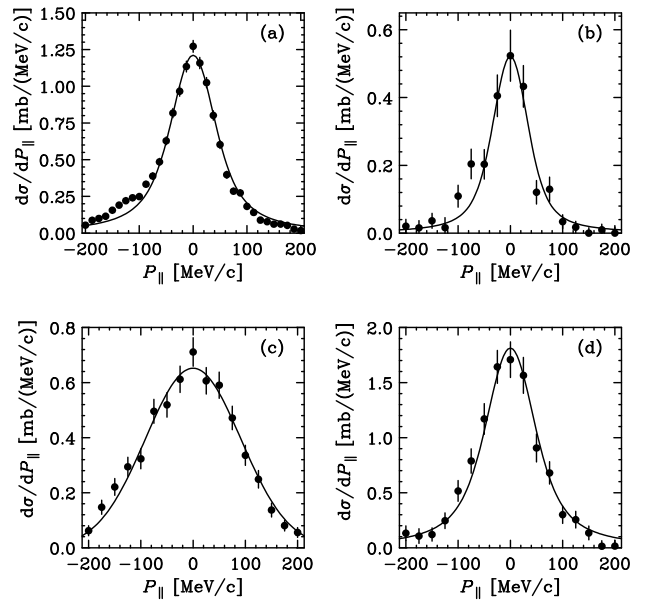}
\caption{The momentum distributions for (a) C($^{19}$C, $^{18}$C), (b)
C($^{20}$C, $^{19}$C), (c) C($^{20}$C, $^{18}$C), and (d) C($^{22}$C,
$^{20}$C). The shapes of the distributions for cases (a), (b), and (d)
are well reproduced by a Lorentzian distribution convoluted with the
experimental resolution. For the ($^{20}$C, $^{18}$C), case (c), the
distribution is better reproduced by a Gaussian. The distribution
shapes were fitted over the momentum ranges $-$200 to $+$200~MeV/$c$.}
\label{fig:fit}
\end{center}
\end{figure}

The measured momentum distributions are shown in Fig.~\ref{fig:fit}.
That for ($^{22}$C, $^{20}$C) was obtained for the first time
here. The shapes of the distributions for ($^{19}$C, $^{18}$C),
($^{20}$C, $^{19}$C), and ($^{22}$C, $^{20}$C) are well reproduced by
a Lorentzian distribution convoluted with the experimental resolutions
of 23, 23, and 27~MeV/$c$, respectively. For ($^{20}$C, $^{18}$C), the
distribution is better reproduced by a Gaussian convoluted with the
experimental resolution of 28~MeV/$c$. The widths (FWHM) after
accounting for the experimental resolutions are listed in Table
\ref{tbl:expcs}.

The distributions for ($^{19}$C, $^{18}$C) and ($^{22}$C, $^{20}$C)
are relatively narrow, with widths of 73(5) and
76(8)~MeV/$c$. Qualitatively, these arise from the weakly-bound one
and two-neutron halo-like nature of these nuclei and the role of
$s$-wave valence neutron(s) in their ground state structure. It is
interesting that the measured distribution for ($^{20}$C, $^{19}$C) is
also extremely narrow. Qualitatively, this is the result of the
reaction probing directly the $\nu{s}_{1/2}^2$ component of the
$^{20}$C($0^+$) ground state wave function, as is required to populate
the bound $^{19}$C$_{\rm g.s.}$($1/2^+$) halo-state residues. The
quantitative analyses for these momentum distributions and their cross
sections are detailed in the following sections.

\section{Reaction Analysis \label{framework}}
\subsection{Reaction theory \label{reaction}}

We adopt an eikonal model description of the reaction mechanisms.
Given a nucleon- or nucleus-target interaction description, the
eikonal approximation has been shown \cite{EB} to provide a rather
accurate description of the elastic $S$-matrix and derived observables
for incident projectile energies of order 50 MeV/nucleon and greater.
As noted earlier, at the energy of the current experiments (about
240~MeV/nucleon) the sudden and eikonal approximations of the reaction
model are very accurate.

The removal reaction cross sections for one-neutron knockout to a
given final state, with spin-parity $J^{\pi}$, are calculated using
\cite{HaT03}
\begin{equation}
\sigma_{-1n}=\sum_{n\ell j}\left[\frac{A}{A-1}\right]^N \!\! \;
C^2S(J^{\pi},n \ell j) \; \sigma_{\rm sp}(n \ell j,S_n^{\rm eff}),
\label{eqn:single_particle}
\end{equation}
where the $C^2S$ are the shell model spectroscopic factors and the
single-particle cross section $\sigma_{\rm sp}$ is calculated using
the eikonal model assuming unit spectroscopic factor. The quantum
numbers of the removed neutron are denoted by $n \ell j$ and $S^{\rm
eff}_n$ is the effective separation energy for the removal of the
neutron leaving the residue in the given final state. The
single-particle cross sections, $\sigma_{\rm sp}$, include the
contributions from both the stripping (inelastic breakup) and
diffractive dissociation (elastic breakup) mechanisms. Details of
calculations of these two distinct and (incoherent) additive
contributions can be found in Ref. \cite{Tos01}.

In direct two-neutron removal the theoretical cross sections do not in
general (e.g., when there are several active orbitals) factorize into
a structural (spectroscopic) factor and a single-particle cross
section. The cross sections involve coherent contributions from all
active shell-model two-nucleon configurations with non-vanishing
two-nucleon amplitudes (TNA). Details of their definitions and the
phase conventions used can be found in Ref.\ \cite{Tos04}. Here we
will calculate the single-step direct two-neutron removal yields
arising from both (a) two-neutron stripping and (b) one neutron being
stripped and the second being elastically dissociated (diffracted)
\cite{Tos06}. Since these direct two-neutron removal cross sections
are small compared to the cross sections arising from single-neutron
removal, we do not expand upon these formal aspects here. Full details
of the necessary eikonal formalism, as is applied to direct
two-nucleon removal events, can be found in Refs. \cite{Tos04,Tos06}.

For both the one- and two-neutron removal calculations, the required
neutron- and residue-target elastic $S$-matrices were calculated using
the static density limit of the eikonal model, e.g., \cite{AlK96},
also known as the $t \rho \rho$ limit of the Glauber multiple
scattering series. That is, we used the single-folding model
(nucleon-target) and double-folding model (residue-target) for the
absorptive optical model interactions with the carbon target. The
inputs needed were the residue and target point neutron and proton
one-body densities and an effective nucleon-nucleon ($NN$)
interaction. The densities of the $^{18,19,20,21}$C residues were
estimated from spherical Skyrme Hartree-Fock (HF) calculations using
the SkX interaction \cite{Bro98}.  The HF calculates the
(experimentally unbound) $^{21}$C case to be weakly bound. We used
this bound density for calculations of the $^{21}$C-target optical
potential (and its $S$-matrix) in the localized region where they make
grazing collisions.

All calculations assumed the following. The density of the carbon
target nuclei was taken to be of Gaussian form with a point-nucleon
root-mean squared radius of 2.32 fm. A zero-range $NN$ effective
interaction was used with its strength calculated from the free
neutron-neutron and neutron-proton cross sections at the beam energy
and from the real-to-imaginary ratios of the $NN$ forward scattering
amplitudes at 240 MeV, interpolated (using a polynomial fit) from the
values tabulated by Ray \cite{Ray79}. The use of these inputs, as a
standard parameter set in the eikonal reaction model, was shown to
accurately reproduce the recently-measured \cite{DB1} ratios of the
diffraction to stripping reaction mechanism yields in the cases of
$^8$B and $^9$C induced proton-removal reactions. A recent careful
analysis by Bertulani and De Conti \cite{Carlos} confirms that
corrections to this adopted procedure, due to Pauli blocking
corrections to the $NN$ effective interaction, are negligible at the
energy of the present study.

We assume here that the heavy residue-target interactions and their
$S$-matrices are diagonal with respect to the different final states
of the residue, and thus that there is no reaction-induced dynamical
excitation of the residues during the collision. For the odd-$A$
carbon projectiles, where different neutron orbitals ($n\ell j$) may
then contribute to a given $J^{\pi}$ final state, this implies that
these different $n\ell j$ contributions are incoherent and should
be summed.

\subsection{Shell-model calculations and overlaps \label{shell}}

A consistent set of shell-model calculations were used for the
required level energies, spectroscopic factors and two-nucleon
amplitudes. These were performed using the code {\sc oxbash}
\cite{oxbash}. The calculations used the Warburton and Brown (WBP)
effective interaction \cite{War92} in a $psd$-model space truncated to
allow 0$\hbar \omega$ and 1$\hbar \omega$ excitations. The small
center of mass correction factor, $\left[ A/ (A-1) \right]^N$, shown
in Eq.\ (\ref{eqn:single_particle}), where $N$ is the principal
oscillator model quantum number of the orbit of the removed nucleon
\cite{Die74}, was applied to the (fixed center) shell-model
spectroscopic factors of all of the single-neutron removal
calculations.

The low-lying shell-model levels and spectroscopic factors for the
reaction products, $^{18,19,20,21}$C, are collected in Tables
\ref{tbl:res1n} and \ref{tbl:res1nn}. The wave function of each of the
removed-neutron bound states was calculated in a Woods-Saxon potential
well of a fixed geometry. Following Ref.\ \cite{ECS_C}, these radial
wave functions were calculated using a standard Woods-Saxon potential
geometry with a reduced radius parameter $r_0$ = 1.25 fm and a
diffuseness of $a_0$ = 0.7 fm.

The depths of the potential wells in each instance were adjusted to
reproduce the neutron separation energy, taking into account the
excitation energy of the final state. The ground state to
ground state one- and two-neutron separation energies were taken
from the 2003 mass evaluation \cite{Aud03}. The shell-model
excitation energies of Tables \ref{tbl:res1n} and \ref{tbl:res1nn}
were then used (without any adjustment) for all non-ground-state
transition calculations. For the direct two-neutron removal
calculations each neutron was assumed to have a separation energy of
half that for the two neutrons to the final state of interest.

\section{Analysis and Discussion \label{results}}

We discuss the results for the inclusive cross sections for one- and
two-neutron removal from the $^{19,20,22}$C isotopes. We consider in
detail the calculated contributions from both indirect and direct
two-nucleon removal. The measured and calculated inclusive parallel
momentum distributions are also discussed. In all cases these are
shown in the projectile rest frame. In the comparisons with the data
the theoretical momentum distributions, calculated using the stripping
mechanism, have been convoluted with the Gaussian experimental
resolution given in Section \ref{exp_results} and then scaled to the
measured inclusive cross
sections. Further discussion of the calculations of the parallel
momentum distributions in the cases of the transitions to unbound
final states is included when discussing these; i.e.,  for the
$^{20,22}$C cases.

\subsection{One-neutron removal reactions}

We first discuss the individual and the inclusive one-neutron removal
cross sections to bound, $\sigma_{-1n}^{\rm th}$, and unbound (neutron
emitting), $\sigma_{-1n(e)}^{\rm th}$, states of the mass $A-1$
isotopes. The experimental and theoretical results of the present work
are collected in Tables \ref{tbl:res1n} and \ref{tbl:res1nn}. Also
tabulated are the details of the relevant shell-model states, their
energies, spins, parities and spectroscopic factors, $C^2S$. The
overall ratios of the measured to the calculated inclusive one-neutron
removal cross sections, $R_s= \sigma_{-1n}^{\rm exp}/\sigma_{-1n}^{\rm
th}$, are also shown in the tables for each bound and unbound final
states data set.

\subsubsection{Results for $^{19}$C\label{19Csection}}

The case of $^{19}$C provides a valuable link to the earlier work at
lower energies, summarized in \cite{ECS_C}, and the related and more
exclusive results using neutron knockout from a proton target
\cite{Kondo}. From the present work, calculated exclusive and
experimental inclusive one-neutron removal yields from the
$^{19}$C($1/2^+$) ground state, with ground-state separation energy
$S_{1n}(^{19}{\rm C})$ = 0.58 MeV, are shown in Table \ref{tbl:res1n}.
The theoretical cross sections are shown for the six predicted
positive parity $^{18}$C final states.

In the case of ($^{19}$C, $^{18}$C) the WBP shell model calculation
places several final states near to or between the one- and
two-neutron threshold energies of 4.18 MeV and 4.91 MeV, respectively,
for $^{18}$C. Specifically, the third $2^+_3$ and first 3$^+_1$ states
at 4.915 and 4.975 MeV have significant spectroscopic strengths and
associated cross sections. Experimentally, recent work of Kondo {\em
et al.}  \cite{Kondo}, on neutron knockout from $^{19}$C on a proton
target, observed gamma-rays from a $(2^+,3^+)$ excited state (or
states) near 4.0 MeV, the associated $^{18}$C transverse momentum
distribution being characteristic of an $\ell=2$ neutron removal. The
earlier ($^{19}$C, $^{18}$C) inclusive data analyses of Maddalena {\em
et al.} \cite{Mad01} and Simpson and Tostevin \cite{ECS_C} also
assumed these 2$^+_3$ and 3$^+_1$ states near 4.9 MeV were neutron
bound, citing the results of shell-model calculations using a modified
version of the WBT interaction \cite{Sta08}.

\begin{figure}[tbp]
\begin{center}
\includegraphics[width=0.95\columnwidth]{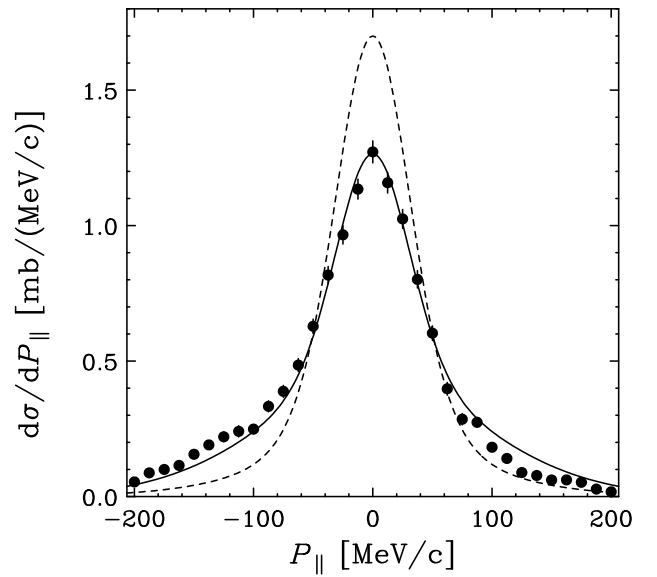}
\end{center}
\caption{Measured inclusive parallel momentum distribution of
$^{18}$C, following one-neutron removal from $^{19}$C on a carbon
target at 243 MeV/nucleon compared to the theoretical
calculations. The solid line includes contributions from the 2$_3^+$
and 3$_1^+$ shell-model states of $^{18}$C, assumed bound; see also
Table \protect\ref{tbl:res1n}. The dashed line shows the results when
assuming that the 2$_3^+$ and 3$_1^+$ states are unbound. Here, and in
Fig.~\ref{fig:20-19C}-\ref{fig:22-20C2}, the theoretical distributions
have been convoluted with the experimental resolution and normalized
to the measured inclusive cross section.}
\label{fig:19-18C}
\end{figure}

We have also calculated the inclusive parallel momentum distributions,
to bound final states, as the sum of the distributions to these
individual final states weighted by the $\sigma_{-1n}^{\rm th}$ shown
in Table~\ref{tbl:res1n}. Fig.~\ref{fig:19-18C} shows the experimental
($^{19}$C, $^{18}$C) inclusive parallel momentum distribution and also
those calculated. In all cases the theoretical momentum distribution
curves are normalised to the measured inclusive cross section. We show
the results obtained by (a) assuming that the 2$_3^+$ and 3$_1^+$
states are unbound (dashed curve), having an inclusive cross section
of 100.2 mb, and (b) assuming that the 2$_3^+$ and 3$_1^+$ states are
bound (solid curve), resulting in an inclusive cross section of 164.4
mb. The experimental cross section, in Table \ref{tbl:res1n}, is
163(12) mb. The comparison with the present momentum distribution
data, in particular, provides us with rather compelling evidence for
the hypothesis (b), that the 2$_3^+$ and 3$_1^+$ states {\em are}
bound.

\subsubsection{Results for $^{20}$C}

The predicted $^{19}$C shell-model final states and the calculated and
experimental one-neutron removal cross sections from the
$^{20}$C($0^+$) ground state, with separation energy
$S_{1n}$($^{20}$C) = 2.90 MeV are collected in Tables \ref{tbl:res1n}
and \ref{tbl:res1nn}. There is only very incomplete experimental
information on the low-lying excited state spectrum of $^{19}$C. Using
Coulomb dissociation the $^{19}$C ground state has been unambiguously
identified as a $1/2^+$ $s$-wave halo state with weak binding
\cite{Naka19}. The evaluated $^{19}$C first neutron threshold is at
0.58(9) MeV \cite{Aud03}. An unbound excited $5/2^+$ state with
$E_{\rm x}$ = 1.46(10) MeV has also been clearly identified
\cite{Satou} using inverse-kinematics proton inelastic scattering from
$^{19}$C. Stanoiu {\em et al.} \cite{Sta08} reported a 201(15) keV
gamma-ray transition in $^{19}$C with in-beam $\gamma$-ray
spectroscopy.  Using inverse-kinematics proton inelastic scattering
Elekes {\em et al.}  \cite{Elk19} also reported evidence of two
gamma-ray transitions, with energies 72(4) and 197(6) keV, from two
bound $^{19}$C excited states.  While in both of these cases the
transition energy (near 200 keV) is close to that of a predicted
$5/2^+$ shell-model bound excited state, we will show that the present
experimental data and analysis do not support such a strong transition
to a $5/2^+$ bound $^{19}$C excited state.

\begin{figure}[tbp]
\begin{center}
\includegraphics[width=0.95\columnwidth]{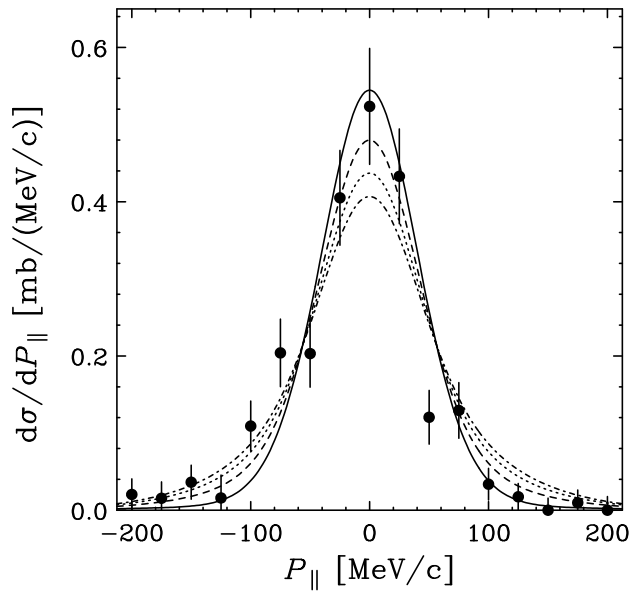}
\end{center}
\caption{Measured inclusive parallel momentum distribution of
$^{19}$C, following one-neutron removal from $^{20}$C on a carbon
target at 241 MeV/nucleon compared to the theoretical
calculations. The solid curve assumes that only the 1/2$^+$
shell-model ground state transition ($2s_{ 1/2}$ neutron removal) is
bound. The long-dashed, short-dashed, and dot-dashed curves result if
one assumes that 0.5, 1.0 or 1.5 units of $1d_{5/2}$ spectroscopic
strength also leads to bound final states.}
\label{fig:20-19C}
\end{figure}

Table \ref{tbl:res1n} shows the cross section for the 1/2$^+_1$
shell-model ground state transition. The measured cross section, of
58(5) mb, and parallel momentum distribution to a bound $^{19}$C final
state, Fig.~\ref{fig:20-19C}, are consistent with the theoretical
expectations for the removal of a $2s_{1/2}$ neutron (solid curve)
with the tabulated shell-model spectroscopic factor of near to
unity. It is likely that the 1/2$^+$ ground state of $^{19}$C is the
only bound state of this system. If any bound excited state exists,
the only possibility seems to be the first 3/2$^+_1$ state predicted
at 0.624 MeV, which would add the cross section of only 7.20 mb.

Although the bound 5/2$^+$ state is unplausible due to the observed
small cross section, we attempt to estimate an upper limit on possible
bound $d$-state strength below. Table \ref{tbl:res1nn} shows the
results for the cross sections leading to the excited $^{19}$C
shell-model final states. The shell model predicts seven such excited
states with significant spectroscopic factors below the $^{18}$C
neutron threshold of 4.18 MeV. Given these cross sections we note,
from Table \ref{tbl:res1nn}, that one unit of the first excited state
$1d_{5/2}$ spectroscopic strength makes a contribution of 30.5 mb to
the theoretical cross section. The calculated 1/2$^+_1$ ground state
cross section is 58.92 mb. Thus, if there was also $1d_{5/2}$ strength
to bound state(s), with a summed spectroscopic strength of 0.5, 1.0,
or 1.5 units, the theoretical cross section to bound final states
would increase to 74.2, 89.4, or 104.7 mb, respectively, well in
excess of the measured value of 58(5) mb. The corresponding effects of
such bound $1d_{5/2}$ strength on the shapes of the calculated
$^{19}$C parallel momentum distributions are shown in
Fig.~\ref{fig:20-19C} by the long-dashed, short-dashed, and dot-dashed
curves. Here, each curve is normalized to the experimental cross
section of 58(5)~mb. We conclude from this comparison that the
majority of the strength that the shell-model attributes to the 190
keV $1d_{5/2}$ state is in fact unbound. Based on
Fig.~\ref{fig:20-19C} and the measured cross section to bound
$^{19}$C, we estimate that 0.5 units or less of bound $1d_{5/2}$
strength might be accommodated by the present data set.

\begin{figure}[tbp]
\begin{center}
\includegraphics[width=0.95\columnwidth]{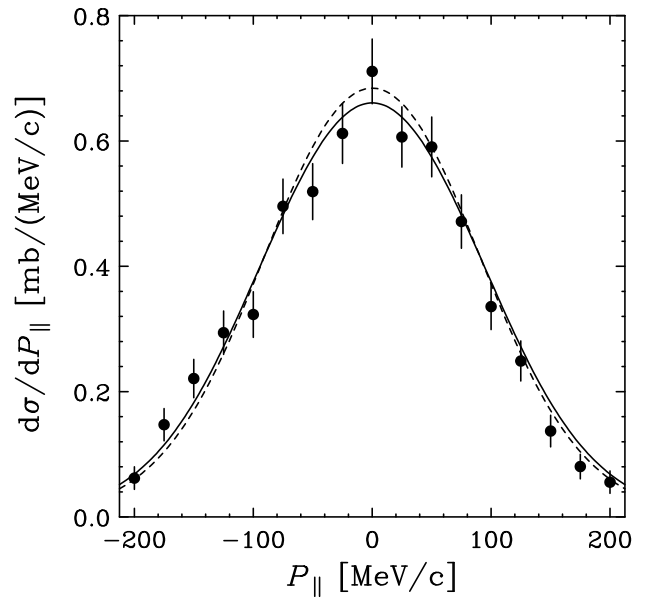}
\end{center}
\caption{Measured inclusive parallel momentum distribution of
$^{18}$C, following two-neutron removal from $^{20}$C on a carbon
target at 241 MeV/nucleon compared to the theoretical
calculations. The theoretical curves are the weighted sum of the
exclusive calculations of the unbound $^{19}$C states, see
text. Recoil effects associated with the neutron emission are included
assuming the most important contributions come from states with
$\varepsilon^*$ of 1.0 MeV (dashed curve) and 2.0 MeV (solid curve).}
\label{fig:20-18C}
\end{figure}

Our assumption, in Table \ref{tbl:res1nn}, is that all of the excited
$^{19}$C shell-model states are unbound and that these unbound states
will decay by neutron emission to $^{18}$C. In this and the following
case of $^{22}$C these unbound mass $A-1$ excited state cross sections
are large. For such unbound final state cases our one-neutron removal
model calculates the exclusive parallel momentum distributions of the
(weakly) unbound $^{19}$C and $^{21}$C residues in the original
projectile rest frame. The subsequent in-flight neutron emission from
these excited states will generate additional (recoil) broadening of
the momentum distributions of the observed mass $A-2$ residues, i.e.,
$^{18}$C and $^{20}$C. The degree of broadening will be dependent on
the continuum energy of the unbound, mass $A-1$ intermediate state,
denoted $\varepsilon^*$.

\begin{figure}[tbp]
\begin{center}
\includegraphics[width=0.95\columnwidth]{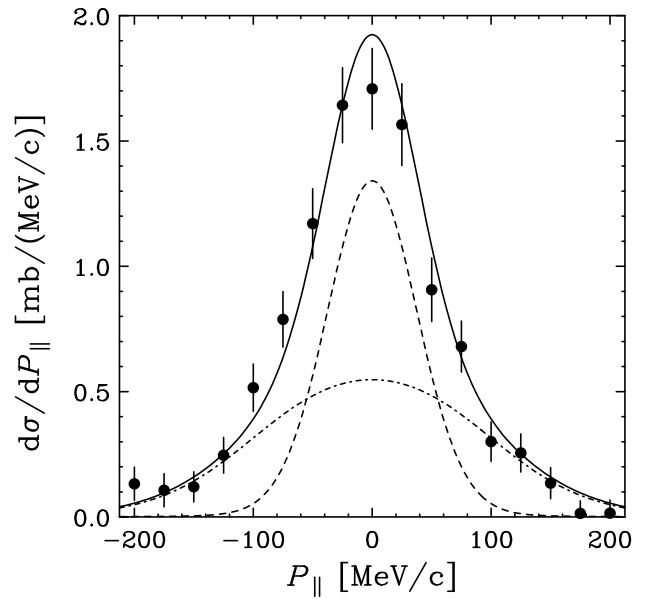}
\end{center}
\caption{Measured inclusive parallel momentum distribution of
$^{20}$C, following two-neutron removal from $^{22}$C on a carbon
target at 240 MeV/nucleon compared to the theoretical
calculations. The solid curve is the weighted sum of the exclusive
calculations for the unbound $^{21}$C states, see text. The dashed and
dot-dashed curves show the contributions from knockout via the
1/2$_1^+$ and 5/2$_1^+$ unbound $^{21}$C intermediate states,
respectively. The recoil broadening arising from neutron emission from
these unbound intermediate states is folded in.}
\label{fig:22-20C}
\end{figure}

We estimate the effect of this recoil. We assume that, in the rest
frame of the unbound, mass $A-1$ state, with its given continuum
energy $\varepsilon^*$, the mass $A-2$ residue (in its ground state)
and the decay neutron separate isotropically. The momentum $p$ of the
heavy decay residue in this frame satisfies $p^2 = 2 \mu
\varepsilon^*$, where $\mu$ is the $A-2$ residue-neutron reduced
mass. The assumption that this two-body decay is isotropic then
requires that the calculated parallel momentum distributions of the
unbound mass $A-1$ fragments must be convoluted with a rectangular
distribution, of unit integral and total width $2p$, to derive the
mass $A-2$ fragment parallel momentum distributions. This is done for
the theoretical distributions shown in Figs.~\ref{fig:20-18C} and
\ref{fig:22-20C}.

Table \ref{tbl:res1nn} shows that these indirect two-neutron removal
cross sections arise predominantly from intermediate states of small
$\varepsilon^*$. For the $^{20}$C case, a cross section of 160 mb is
predicted to arise from the first three shell-model excited states
having $E_{\rm x} < 1$ MeV in $^{19}$C. However, as was discussed
above, these shell model energies are not sufficiently accurate. The
WBP interaction predicts the first excited $5/2_1^+$ state to be bound
with a large spectroscopic factor, whereas the calculated
$\sigma_{-1n}^{\rm th}$ to this state and the measured cross section
and momentum distribution to the $^{19}$C ground state exclude this
possibility. Experimentally a $^{19}$C excited $5/2^+$ state has been
clearly identified at $E_{\rm x}$ = 1.46(10) MeV by Satou {\em et al.}
\cite{Satou} in proton inelastic scattering from $^{19}$C, i.e., with
$\varepsilon^*$ = 0.88 MeV. The present inclusive data do not permit a
more detailed analysis of this excited state or of the predicted
shell-model strength distributions. Our analysis shows, however, that
the present data are consistent with an integrated $5/2^+$ strength of
about 4 units leading to the $^{19}$C continuum, as is given by the
shell model.

In the absence of more complete information, and to assess this recoil
sensitivity, we calculate the evaporation recoil effects assuming that
the most important contributions arise from $^{19}$C intermediate
states with (i) $\varepsilon^*$ = 1.0 MeV and (ii) $\varepsilon^*$ =
2.0 MeV. The results are shown in Fig.~\ref{fig:20-18C} by the dashed
line and solid line, respectively.  Both outcomes are consistent with
the experimental $^{18}$C residue momentum distribution. We conclude
from this agreement of the inclusive cross section and the momentum
distribution that the $5/2^+_1$ shell-model excited state is very
probably unbound. Since, in this case, the shell model appears to
systematically produce states with too small an excitation energy, the
effective neutron separation energies will also be underestimated and,
in turn, the theoretical removal cross sections slightly
overestimated. This is also suggested by the smaller $R_s=
\sigma_{-2n}^{\rm exp}/ \sigma_{ -1n(e)}^{\rm th}$ deduced for this
isotope. We do not attempt to make any parameter adjustments to
compensate for this (small) effect.

\subsubsection{Results for $^{22}$C}

Here all final states of the $^{21}$C one-neutron removal residues are
particle unbound. The calculated exclusive (and inclusive) and
experimental inclusive one-neutron removal yields are collected in the
Table \ref{tbl:res1nn} for the predicted shell-model states of
$^{21}$C, that decay by neutron emission to $^{20}$C.

Very little is known about these isotopes. Both the one- and
two-neutron separation energies from $^{22}$C are only poorly
determined and so we are guided by the 2003 mass evaluation
\cite{Aud03}. That is $S_{2n}$($^{22}$C) = 0.42(94) MeV and
$S_{1n}$($^{21}$C) = $-0.33(56)$ MeV, with large uncertainties. Thus,
the ground state of $^{21}$C was assumed to be produced at a continuum
energy of $\varepsilon^*$ = 0.30 MeV after neutron removal with ground
state separation energy $S_{1n}$($^{22}$C) = 0.70 MeV. As was
discussed for the $^{20}$C projectile case, the inclusive (unbound)
$^{21}$C momentum distribution is calculated as the weighted sum of
the momentum distributions to the individual final states with the
$\sigma^{\rm th}_{-1n(e)}$ shown in Table \ref{tbl:res1nn}.  The
neutron emission recoil broadening is included for each final state
according to its $\varepsilon^*$ value, i.e., $\varepsilon^* = E_{\rm
x}+0.30$ MeV, prior to this sum being performed. Three final states
are predicted below the $^{20}$C first neutron threshold of 2.90 MeV.

\begin{figure}[tbp]
\begin{center}
\includegraphics[width=0.95\columnwidth]{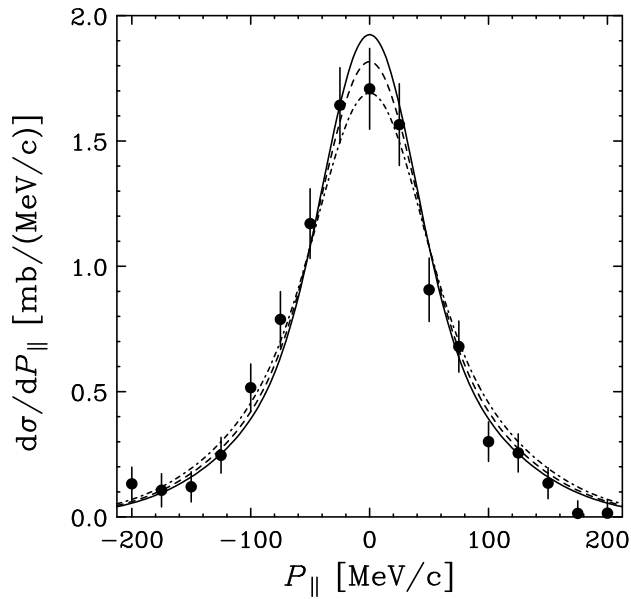}
\end{center}
\caption{Measured inclusive parallel momentum distribution of
$^{20}$C, following two-neutron removal from $^{22}$C on a carbon
target at 240 MeV/nucleon compared to the theoretical
calculations. The theoretical curves are the inclusive cross sections
calculated assuming $^{22}$C two-neutron separation energies
$S_{2n}$($^{22}$C) = 0.40~MeV (solid curve), 0.70~MeV (dashed curve),
and 1.20~MeV (dot-dashed curve). The curves include the recoil
broadening arising from the neutron decay of the unbound $^{21}$C
intermediate states.}
\label{fig:22-20C2}
\end{figure}

Table \ref{tbl:res1nn} shows that, based on the shell-model, the first
two final states each contribute almost half of the inclusive
one-neutron removal cross section. These states are a $1/2_1^+$ ground
state, with spectroscopic factor $C^2S$ = 1.4, and a $5/2_1^+$
neutron-hole state at $E_{\rm x}$ = 1.11~MeV, with $C^2S$ = 4.2. The
associated measured and theoretical inclusive $^{20}$C parallel
momentum distributions (convoluted with the experimental resolution of
27~MeV/$c$) are compared in Fig.~\ref{fig:22-20C}. The individual
contributions from the two dominant shell-model final states are also
shown in the figure. The agreement with the data is very good,
providing strong support for the weakly-bound, $\nu{s}_{1/2}^2$
character for the $^{22}$C ground state. This result is consistent
with the recent interaction cross section measurement and associated
analysis of Ref.\ \cite{Tan10}, that is suggestive of an extended
$^{22}$C matter density.

Currently, the $^{22}$C two-neutron separation energy,
$S_{2n}$($^{22}$C) = 0.42(94) MeV, has a significant
uncertainty. Hence, we consider the sensitivity of the theoretical
inclusive $^{20}$C production cross section and momentum distribution
to the value assumed. Fig.~\ref{fig:22-20C2} shows the calculated
momentum distributions when assuming a $^{22}$C two-neutron separation
energy of $S_{2n}$($^{22}$C) = 0.40~MeV (solid curve), 0.70~MeV
(dashed curve), and 1.20~MeV (dot-dashed curve). In these calculations
we continue to assume that the ground state of $^{21}$C is at a
continuum energy of 0.30~MeV, hence the ground state to ground state
one-neutron separation energy of $^{22}$C is $S_{1n}$($^{22}$C) =
0.70, 1.00, and 1.50~MeV in these cases.  The curves have been
convoluted with the experimental resolution of 27~MeV/$c$ and also
include the recoil broadening arising from the neutron decay of the
unbound $^{21}$C states (Table \ref{tbl:res1nn}). The increasing
separation energies reduce the corresponding inclusive cross sections:
283, 257, and 227~mb for the $S_{2n} $($^{22}$C) = 0.40, 0.70, and
1.20~MeV, respectively. We note that, owing to the relative
insensitivity of our calculated cross sections and momentum
distributions to the current experimental uncertainty in
$S_{2n}$($^{22}$C), the data of the present work do not determine or
place a significant constraint upon this value. We are able to
conclude, however, that the sensitivity to the underlying structure of
$^{22}$C, specifically of the approximately equal contributions of the
$1/2_1^+$ and $5/2_1^+$ transitions to the measured inclusive cross
section, is robust.

\subsection{Direct two-neutron removal}

We summarize only briefly the calculated exclusive and inclusive {\em
direct} two-neutron removal cross sections, $\sigma_{-2n(d)}^{\rm
th}$, from $^{20}$C and $^{22}$C to bound states of the mass $A-2$
isotopes. These results are collected in Table \ref{tbl:res2nn},
computed based on the WBP shell-model two nucleon amplitudes, TNA. As
noted earlier, as these calculated cross sections were both expected
and found to be small, we will not enter into an extended discussion
and details of the calculations. The descriptions of the nucleon
overlap functions used and the construction of the residue- and
neutron-target $S$-matrices are the same as for the one-neutron
removal analysis. For the full details and the formalism of the
exclusive cross sections (and their momentum distributions) the reader
is referred to recent references \cite{Tos06,simxx}.

For $^{20}$C projectiles six states below the neutron threshold in
$^{18}$C have appreciable TNA. These states include the 2$_3^+$ and
3$_1^+$ states proposed as being bound from the one-neutron removal
analysis (Section \ref{19Csection}). The inclusive direct two-neutron
removal cross section is calculated to be 14.6 mb. For $^{22}$C
projectiles, just two states below the neutron threshold in $^{20}$C
have appreciable TNA and the direct two-neutron removal cross section
is now 12.1 mb. These numbers are to be compared with those for the
indirect two-neutron removal paths that predict cross sections of
191.2 and 283.0 mb, respectively. In addition we note that, in the
case of the removal of strongly-bound two-neutron pairs, these
calculated direct two-nucleon removal cross sections typically
overestimate the measured cross sections with
$R_s(2N)=\sigma_{-2n}^{\rm exp} / \sigma_{-2n(d)}^{\rm th}\approx 0.5$
\cite{Tos06}. Thus, as was found in the earlier study of the lighter
carbon isotopes \cite{ECS_C}, the direct pathways enter at about an
8\% level. Since we are unable to distinguish these direct events with
the current experimental setup, they cannot be elucidated or exploited
further here.

\section{Conclusions}

In summary, the structure of the most neutron-rich carbon isotopes
$^{19,20,22}$C has been investigated using high-energy (about
240~MeV/nucleon) single and two-neutron removal reactions on a carbon
target. Narrow momentum distributions were observed for one-neutron
removal from $^{19}$C and $^{20}$C and two-neutron removal from
$^{22}$C. A much broader momentum distribution was found in the case
of two-neutron removal from $^{20}$C.

The measured cross sections and momentum distributions were
interpreted in the light of eikonal reaction model calculations for
single-neutron knockout combined with structural input derived from
$psd$ shell-model calculations employing the WBP interaction. The
two-neutron removal cross sections were calculated by considering (a)
the removal of one neutron to unbound states in the $A-1$ daughter,
with the assumption that these unbound intermediate states decay by
neutron emission to bound states in the mass $A-2$ residue and (b)
direct two-neutron removal.

In the case of C($^{22}$C,$^{20}$C), the cross section and momentum
distribution are consistent with the existence of a two-neutron halo
with a dominant $\nu{s}_{1/2}^2$ configuration in $^{22}$C. The very
narrow momentum distribution and relatively low cross section for
C($^{20}$C,$^{19}$C) arises as the single-neutron removal to the
$^{19}$C(1/2$^+$) ground state probes specifically the significant
$\nu{s}_{1/2}^2$ component of the $^{20}$C ground state. The
$\nu{d}_{5/2}$ component results in the population of unbound states
in $^{19}$C that neutron decay to $^{18}$C. The narrow momentum
distribution and enhanced cross section for C($^{19}$C,$^{18}$C) are
consistent with the well developed $\nu{s}_{1/2}$ halo of $^{19}$C.

Overall, the calculated cross sections agreed well with those
measured.  In particular, in the cases of C($^{19}$C,$^{18}$C),
C($^{20}$C,$^{19}$C), and C($^{22}$C,$^{20}$C), $R_s = \sigma^{\rm
exp}_{-1n} / ^{\rm th}_{-1n}$ was close to unity and consistent with
systematics \cite{Gad08}. Combined with the good agreement for the
momentum distributions, it may be seen that the shell-model has
predictive power in this region and provides a good overall
description of level positions and their spectroscopic strengths.

In the case of weakly-bound two-neutron removal, fully-correlated,
direct removal cross sections were also calculated. It was shown that
two-neutron removal is dominated by the two-step process; that is,
one-neutron removal followed by neutron decay of the unbound
intermediate state(s). In the cases of two-neutron removal from
$^{20,22}$C, the direct removal contribution was computed to be of
order 8\% or less of the two-step process; consistent with an earlier
work \cite{ECS_C}. Although this makes the identification of these
direct two-nucleon removal events challenging, kinematically complete
measurements of $^{20,22}$C breakup are expected to be possible in the
near future and will help clarify and quantify this process.

\begin{acknowledgments}
We would like to thank the RIKEN accelerator staff and BigRIPS team
for the excellent beam delivery. This work was supported by the Global
Center of Excellence Program ``Nanoscience and Quantum Physics", the
United Kingdom Science and Technology Facilities Council (STFC) under
Grants ST/F012012 and ST/J000051, and NRF grant
(No.~R32-2008-000-10155-0 (WCU)) of MEST Korea. N.K. thankfully
acknowledges the Grant-in-Aid for JSPS Fellows (No.~22$\cdot$9675),
and T.N. the Grant-in-Aid for Scientific Research (B) (No.~22340053)
from the Ministry of Education, Culture, Sports, Science and
Technology, Japan. J.A.T. gratefully acknowledges the financial and
facilities support of the Department of Physics, Tokyo Institute of
Technology, and E.C.S. the support of the United Kingdom Engineering
and Physical Sciences Research Council (EPSRC) under Grant
No.~EP/P503892/1. J.G. and N.A.O. acknowledge partial support from the
Franco-Japanese Nuclear Structure Problems LIA.
\end{acknowledgments}

\bibliographystyle{apsrev}

\begin{widetext}
\clearpage
\begin{table*}[t]
\caption{Results for one-neutron removal reactions from $^{19,20}$C.
Tabulated are the one-neutron removal cross sections to assumed bound
shell-model states near and below the neutron thresholds in the mass
${A-1}$ systems, $^{18,19}$C, of 4.18 and 0.58 MeV, respectively (see
also the footnotes). The final theoretical cross sections,
$\sigma_{-1n}^{\rm th}$, include the center-of-mass correction factor
$\left[A/(A-1\right)]^N$. The errors shown for the ratio of cross
sections, $R_s=\sigma_{-1n}^{\rm exp}/ \sigma_{-1n}^{\rm th}$, reflect
only the errors quoted on the measurements. \label{tbl:res1n} }
\begin{ruledtabular}
\begin{tabular}{lcccccccc}
Reaction \ \ \ & $E_{\rm x}$(MeV)  &  \ \  $J^{\pi}$ \ \  & \ \ $\ell$\ \
& \ \ $\sigma_{\rm sp}$ (mb)  & \ \ $C^2S$\ \ & \ \ $\sigma_{-1n}^{\rm th}$
(mb)   &   \ \  $\sigma_{-1n}^{\rm exp}$
(mb) \ \   & \ \  $R_s$\ \\
\hline\\
($^{19}$C(1/2$^+$),$^{18}$C($J^{\pi}$))              & 0.000 & 0$_1^+$ & 0 & 104.7 & 0.580 & 67.63  &  &  \\
$S_{1n}$($^{19}$C) = 0.58 MeV    & 2.114 & 2$_1^+$ & 2 & 29.91 & 0.470 & 15.67  &  &  \\
                                 & 3.639 & 2$_2^+$ & 2 & 25.91 & 0.104 & 3.00   &  &  \\
                                 & 3.988 & 0$_2^+$ & 0 & 39.35 & 0.319 & 13.97  &  &  \\
                                 & 4.915$^a$ & 3$_1^+$ & 2 & 23.60 & 1.523 & 40.04  &  &  \\
                                 & 4.975$^a$ & 2$_3^+$ & 2 & 23.50 & 0.922 & 24.15 &  &
                                 \\
                                 &       &  Inclusive   &   &      &       & 164.5 & 163(12)  &
                                 0.99(7)\\
\\ ($^{20}$C(0$^+$),$^{19}$C($J^{\pi}$))              & 0.000$^b$ &
1/2$_1^+$ & 0 &48.37 & 1.099 &  58.92  &  &
\\
$S_{1n}$($^{20}$C) = 2.90 MeV        &       & Inclusive  &   & &       &
58.92 &
58(5)  &  0.98(8)  \\
\end{tabular}
\footnotetext[1]{The $^{18}$C 2$_3^+$ and 3$_1^+$ states at 4.915 and
  4.975 MeV are assumed to be bound (see Section
  \protect\ref{19Csection}).}

\footnotetext[2]{There is no evidence from the present work that the
$E_{\rm x}$ = 0.190 MeV, $5/2_1^+$ shell-model state in $^{19}$C is
bound.  This state is included in Table \protect\ref{tbl:res1nn} and
is treated as unbound.}
\end{ruledtabular}
\end{table*}

\bigskip
\begin{table*}[t]
\caption{Results for the {\em indirect} two-neutron removal reaction
cross sections. Tabulated are the one-neutron removal cross sections
to all predicted {\em unbound} ${A-1}$-body shell-model states with
energies below the neutron threshold of the mass ${A-2}$ systems.
That is, the neutron-unbound final states of the intermediate, mass
${A-1}$ systems $^{19,21}$C, below 4.18 and 2.90 MeV,
respectively. The final theoretical cross sections,
$\sigma_{-1n(e)}^{\rm th}$, include the center of mass correction
factor $\left[A/(A-1\right)]^N$. The errors shown on the ratio of
cross sections, $R_s=\sigma_{-2n}^{\rm exp} / \sigma_{-1n(e)}^{\rm
th}$, reflect only the errors quoted on the measurements. The
$\sigma_{-1n(e)}^{\rm th}$ values do not include {\em direct}
two-neutron breakup events, and the $R_s$ values represent upper
limits. \label{tbl:res1nn} }
\begin{ruledtabular}
\begin{tabular}{lcccccccc}
Reaction \ \ \ & $E_{\rm x}$(MeV)  &  \ \  $J^{\pi}$ \ \  & \ \ $\ell$\ \
& \ \ $\sigma_{\rm sp}$ (mb)  & \ \ $C^2S$\ \ & \ \
$\sigma_{-1n(e)}^{\rm th}$ (mb)   &   \ \  $\sigma_{-2n}^{\rm exp}$
(mb) \ \   & \ \  $R_s$\ \\
\hline\\
($^{20}$C(0$^+$),$^{19}$C($J^{\pi}$))              & 0.190$^a$ & 5/2$_1^+$ & 2 & 27.50 & 3.649 & 111.17  &  &  \\
$S_{1n}$($^{20}$C) = 2.90 MeV        & 0.624 & 3/2$_1^+$ & 2 & 26.34 & 0.247 & 7.20    &  &  \\
                                 & 0.927 & 1/2$_1^-$ & 1 & 26.46 & 1.426 & 41.79   &  &  \\
                                 & 1.541 & 5/2$_2^+$ & 2 & 24.31 & 0.282 & 7.59    &  &  \\
                                 & 2.417 & 3/2$_1^-$ & 1 & 22.27 & 0.689 & 17.00   &  &  \\
                                 & 3.284 & 3/2$_2^+$ & 2 & 21.50 & 0.191 & 4.56    &  &  \\
                                 & 3.717 & 1/2$_2^+$ & 0 & 30.53 & 0.055 & 1.86    &  &
                                 \\
                                 &       &  Inclusive   &   &      &       & 191.2 & 155(25)  &
                                 $<$ 0.81(13)\\\\
($^{22}$C(0$^+$),$^{21}$C($J^{\pi}$))              & 0.000 & 1/2$_1^+$ & 0 & 89.35 & 1.403 & 137.55  &  &  \\
$S_{1n}$($^{22}$C) = 0.70 MeV        & 1.109 & 5/2$_1^+$ & 2 & 29.39 & 4.212 & 135.87  &  &  \\
                                 & 2.191 & 3/2$_1^+$ & 2 & 25.44 & 0.342 & 9.55    &  &
                                 \\
                                 &       & Inclusive  &   &           &       & 283.0 & 266(19)  & $<$ 0.94(7)  \\
\end{tabular}
\footnotetext[1]{There is no evidence from the present work that
 $5/2_1^+$ shell-model state in $^{19}$C is
bound. It is assumed to be unbound.}
\end{ruledtabular}
\end{table*}

\begin{table*}[t]
\caption{Theoretical results for the {\em direct} two-neutron removal
reaction cross sections, $\sigma_{-2n(d)}^{\rm th}$. Tabulated are the
two-neutron removal cross sections to all predicted shell-model states
below the neutron threshold in the mass ${A-2}$ systems, $^{18,20}$C
(4.18 and 2.90 MeV, respectively). \label{tbl:res2nn} }
\begin{ruledtabular}
\begin{tabular}{lccc}
Reaction \ \ \ & $E_{\rm x}$(MeV)  & \ \ $J^{\pi}$\ \ & \ \ $\sigma_{-2n(d)}^{\rm th}$ (mb)   \\
\hline\\
($^{20}$C(0$^+$),$^{18}$C($J^{\pi}$)) & 0.000 & 0$_1^+$ & 5.66\\
$S_{2n}$($^{20}$C) = 3.51 MeV                 & 2.114 & 2$_1^+$ & 4.00\\
                                      & 3.639 & 2$_2^+$ & 0.53\\
                                      & 3.988 & 0$_2^+$ &0.36\\
                                      & 4.915$^a$ & 3$_1^+$ &1.98\\
                                      & 4.975$^a$ & 2$_3^+$ &2.10\\
                                      &       &  Inclusive   & 14.6 \\
\\
($^{22}$C(0$^+$),$^{20}$C($J^{\pi}$)) & 0.000 & 0$_1^+$ & 5.32\\
$S_{2n}$($^{22}$C) = 0.40 MeV             & 2.102 & 2$_1^+$ & 6.81\\
                                      &       &  Inclusive  & 12.1 \\
\end{tabular}
\footnotetext[1]{The $^{18}$C 2$_3^+$ and 3$_1^+$ states at 4.915 and
4.975 MeV are assumed to be bound (see Section \protect\ref{19Csection}).}
\end{ruledtabular}
\end{table*}
\end{widetext}

\end{document}